\documentclass[twocolumn,amsmath,amssymb,aps,prl,reprint,longbibliography]{revtex4-1}
\usepackage{graphicx}
\begin{document}

\title{
Active Matter Commensuration and Frustration Effects on Periodic Substrates 
} 
\author{
C. Reichhardt and C. J. O. Reichhardt 
} 
\affiliation{
Theoretical Division and Center for Nonlinear Studies,
Los Alamos National Laboratory, Los Alamos, New Mexico 87545, USA\\ 
} 

\date{\today}
\begin{abstract}
We show that self-driven particles coupled to a periodic obstacle array 
exhibit novel active matter commensuration effects that are absent in
the Brownian limit. As the obstacle size is varied for sufficiently large
activity,
a series of commensuration effects appear
in which the motility induced phase separation
produces commensurate crystalline states, 
while for other obstacle sizes we find frustrated or amorphous states.
The commensuration effects are associated with peaks in the amount of six-fold
ordering and the maximum cluster size.
When a drift force is added to the system,
the mobility contains peaks and dips similar to those
found in transport studies for commensuration
effects in superconducting vortices and colloidal particles.
\end{abstract}
\maketitle

Commensuration effects arise in a variety
of hard and soft matter systems when an assembly
of particles is coupled to a periodic substrate
with a spacing that matches the average interparticle spacing.
Such effects occur for the ordering of atoms or molecules on surfaces
\cite{Bak82,Coppersmith82,Woods14}, 
vortices in superconductors or Bose-Einstein
condensates with periodic pinning arrays
\cite{Harada96,Reichhardt98a,Berdiyorov06,Tung06},
colloidal particles on optical trap arrays
\cite{Brunner02,Reichhardt05,Brazda18}
or patterned surfaces \cite{OrtizAmbriz16}, and cold atoms
on optical lattices \cite{Bloch05}.
Conversely, if the particle assembly 
cannot fit within the constraints imposed by the substrate,
then frustration
can cause the disordering of the system or the formation of
localized defects such as kinks or anti-kinks \cite{Bohlein12,Vanossi12}. 
Commensuration effects
also strongly modify the transport properties
under an applied drive in these systems,
producing
reduced
transport or enhanced pinning
when a commensuration occurs and 
generating a series of
peaks or dips in the transport coefficients as the parameters are varied
\cite{Reichhardt98a,Berdiyorov06,Bohlein12,Vanossi12,Reichhardt17,McDermott13a,Baert95}. 

Coupling of active matter or self-driven particles
to a
substrate
\cite{Marchetti13,Bechinger16}
has been realized in numerous experiments
\cite{Marchetti13,Bechinger16,Palacci13,Buttinoni13,Morin17}. 
Many active particles have only short range repulsive interactions,
so the system forms a uniform liquid at lower densities
in the
non-active or Brownian limit;
however, when activity is present,
the particles undergo
a self-clustering or motility 
induced phase separation into a
high density crystalline phase surrounded by a low density gas
\cite{Palacci13,Buttinoni13,Fily12,Redner13,Cates15}.
Although there
have been various methods proposed for coupling an active matter system
to random
\cite{Bechinger16,Morin17,Reichhardt15,Reichhardt14,Morin17a,Zeitz17,Bhattacharjee19,Chepizhko19,Chardac20} or
periodic obstacle arrays
\cite{Volpe11,AlonsoMatilla19,Ribeiro20,BrunCosmeBruny20,Yazdi20,Reichhardt20},
the possible commensuration effects that could occur on a periodic substrate
in active systems have not been considered
before now.
For a two-dimensional (2D) system of disks in the Brownian or zero 
activity regime, commensuration effects do not
arise
until the disk density $\phi$ is high enough for all of the disks to touch
each other,
so for a nonactive system at $\phi < 0.8$,
commensuration effects should be absent.
Additionally, since thermal 
effects typically wash out commensuration effects \cite{Reichhardt17},
it might be expected that active matter systems 
would not exhibit commensuration effects. 

Here we examine a 2D active matter system of
self-propelled run-and-tumble disks interacting
with a square array of obstacles.
For certain obstacle sizes, we find
that the system can undergo a strong motility-induced phase
transition into a crystalline state that is commensurate with the obstacle
lattice and that coexists with a low density gas.
For other obstacle sizes, the motility-induced phase separation produces
an amorphous crystal
due to a frustration effect caused by a mismatch between the
active disk spacing and the obstacle spacing.
The spacing of the
disks in the
motility-induced dense phase is key in determining whether
commensurate or incommensurate behavior occurs.
The commensuration effects produce peaks in the
size of the largest cluster and in the amount of sixfold ordering.
A variety of different commensurate states
appear,
including states with local square ordering, aligned states, and sliding
crystalline states.  
Under an applied drift force, the transport is a
strongly non-monotonic function of the obstacle size and exhibits
dips at commensurate states as well as
peaks at incommensurate or frustrated states.  
These commensuration effects are absent in the Brownian limit and
become stronger for increasing activity or longer
run times.

{\it Simulation and System---} 
We model a 2D system of active run-and-tumble disks of density $\phi_a$
interacting with a periodic array of obstacles composed of
posts of diameter $d$ and lattice constant $a$.
The overdamped equation of motion
for an active disk $i$ is given by
\begin{equation} 
\alpha_d {\bf v}_{i}  =
{\bf F}^{dd}_{i} + {\bf F}^{m}_{i} + {\bf F}^{obs}_{i} + {\bf F}^{D}_{i} ,
\end{equation}
where the damping constant $\alpha_{d} = 1.0$, and
${\bf r}_i$ and ${\bf v}_i=d{\bf r}_i/dt$ are the position and
velocity of disk $i$.
For the disk-disk interaction force 
${\bf F}^{dd}_{i}$, we use a harmonic repulsion with spring constant $k_{a}$ and 
disk radius $r_{a}$, so that the disk diameter is $d_a=2r_a$.
We set $k_{a} = 150$, which is large enough to keep
the disk-disk overlap in our study below one percent.
The disk-obstacle force ${\bf F}^{obs}$   
is also modeled as a harmonic potential.
The active disk coverage is $\phi_a$ and the combined coverage of the
disks and obstacles is $\phi_{\rm tot}$.  
For the self-propulsion of the active disks ${\bf F}^{m}$,
a force $F_{M}$ is applied
in a randomly chosen direction for a run time of $\tau_{l}$, after which  
the motor force instantaneously reorients to a new
randomly chosen direction for the next run time.
We characterize the
system by the run length $r_l=F_m\tau_l$, the distance
an isolated active particle would move during the run time $\tau_l$.
We also consider the effects of 
an external drive ${\bf F}^D=F_D{\bf \hat x}$
and measure the mobility $M$ using the average velocity in the driving
direction,
$\langle V_{x}\rangle = \sum^{N^{d}}_{i= 1}{\bf v}\cdot {\bf x}$.
We define $M=\langle V_x\rangle/V_{\rm free}$ where
$V_{\rm free}$ is the average velocity that would appear
under the same driving force in the absence of
any obstacles.
We fix $a = 3.0$ and $F_{D} = 0.2$, and
vary
$d_a$, $d$, and $r_{l}$.

\begin{figure}
\includegraphics[width=\columnwidth]{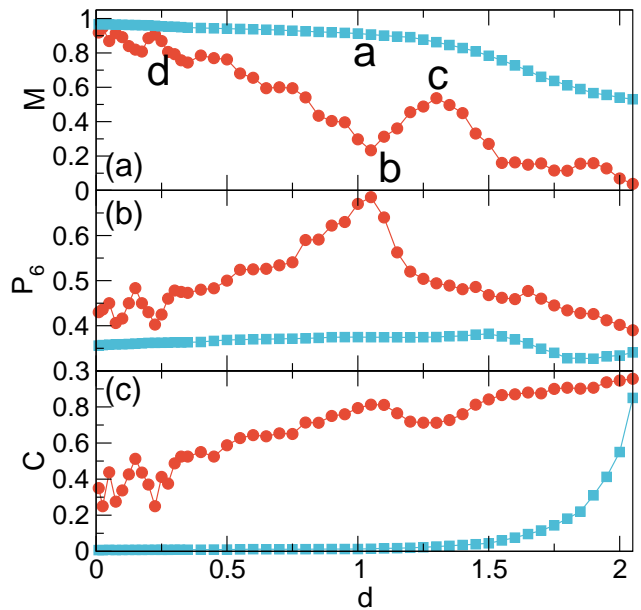}
\caption{ 
Behavior of an active disk system with $d_{a}=0.9$
for a run length of $l_{r} = 0.025$ (squares) in the
Brownian limit and $l_{r} = 175$ (circles) in the active limit.
(a) Mobility $M$ vs obstacle diameter $d$. 
(b) Fraction of six-fold coordinated particles $P_6$ vs $d$.
(c) Size of the largest cluster $C$ vs $d$. 
The letters a-d in panel (a) indicate the points corresponding to the images in
Fig.~\ref{fig:2}.
}
\label{fig:1}
\end{figure}

{\it Results-} 
In Fig.~\ref{fig:1}(a,b,c) we plot the mobility $M$,
the fraction of six-fold coordinated particles $P_{6}$,
and
the fraction of particles in the largest cluster
$C$ versus the obstacle diameter $d$
for a system with $\phi_{a} = 0.32$
and $d_{a} =  0.9$. 
We show two run length values: $l_{r}= 0.025$, where the system is in the
Brownian limit, and
$l_{r} = 175$, the active limit where an obstacle free system would
exhibit motility induced phase separation.
For the short run length of $l_r=0.025$, $M$ has an initial value
near $1.0$ and exhibits a monotonic decrease with increasing $d$,
while $P_{6}$ is mostly flat and $C$ starts to increase once
$d > 1.5$.
For the active limit of $l_r=175$,
$M$ also has an initial value 
near $1.0$ but changes nonmonotonically
with increasing $d$, showing
a pronounced dip near $d = 1.05$ which 
correlates with a peak in $P_{6}$ and
a smaller peak in $C$.
There is also a peak in 
$M$ near $d= 1.3$ that is associated with
a drop in $P_{6}$ and a smaller dip in $C$. Additional
features include a  peak in $M$ near $d = 0.2$
and a smaller peak near $d = 1.75$. 

\begin{figure}
\includegraphics[width=\columnwidth]{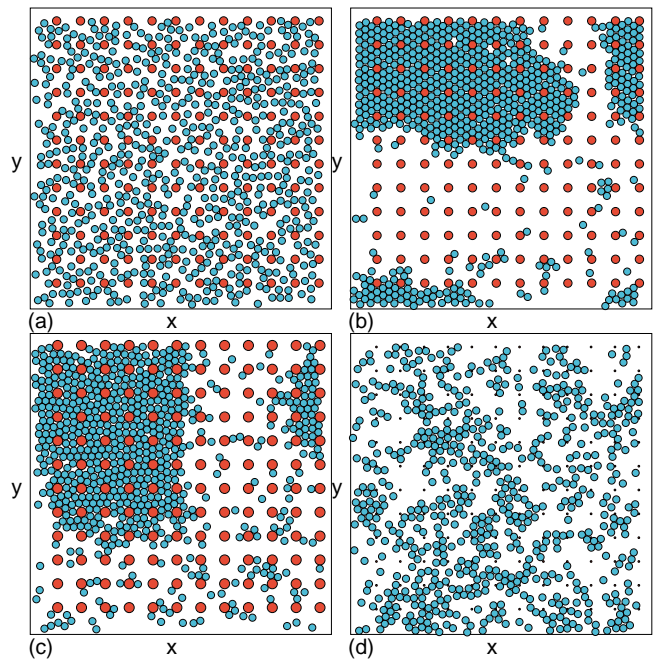}
\caption{ 
Snapshots of the active disk positions (blue circles) and
the obstacles (red circles) for the system in Fig.~\ref{fig:1} with $d_a=0.9$.
(a) A uniform liquid at $d = 1.05$ and $l_{r} = 0.025$.
(b) At $d = 1.05$ and $l_{r} = 175$, there is a phase separated
state in which the dense regions form a commensurate solid.
(c) At $d = 1.3$ and $l_{r} = 175$, there is a phase separated amorphous
or frustrated state.
(d) A frustrated state at $d = 0.225$ and $l_{r} = 175$.
}
\label{fig:2}
\end{figure}

\begin{figure}
\includegraphics[width=\columnwidth]{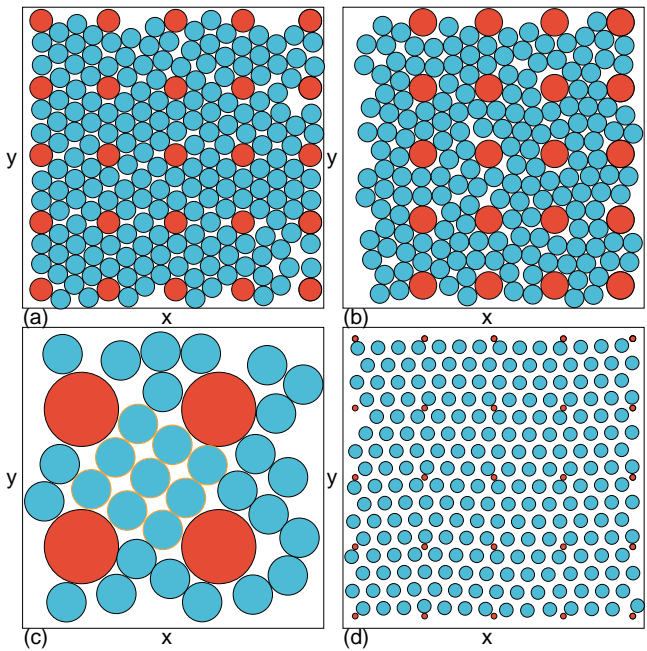}
\caption{Snapshots of the active disk positions (blue circles) and the
obstacles (red circles) for the system in Fig.~\ref{fig:1} with $d_a=0.9$.
(a) A blow up of the dense region in Fig.~\ref{fig:2}(b)
at $d = 1.05$ and $l_{r} = 175$,
showing triangular commensurate ordering.
(b) A blow up of the dense region in Fig.~\ref{fig:2}(c)
at $d = 1.3$ and $l_{r} = 175$, which is in an amorphous state.  
(c) A blow up illustrating the
local square ordering at $d=1.7$ and $l_{r} = 175$.
(d) The full system at $d = 0.15$ and $l_{r} = 175$
showing a sliding crystal phase. 
For clarity, the size of the mobile disks
has been reduced in panel (d).
}
\label{fig:3}
\end{figure}

In Fig.~\ref{fig:2}(a) we show a snapshot of the active particles
and the obstacles for the system in Fig.~\ref{fig:1} 
at $l_{r} = 0.025$ and
$d = 1.05$, where a uniform liquid state appears.
By comparison, in Fig.~\ref{fig:2}(b) a sample with
$l_{r} = 175$  at  $d = 1.05$ forms a phase separated state of
high density coexisting with a low density gas.
This combination of parameters
corresponds to the peak in $P_{6}$ and the dip in $M$ 
in Fig.~\ref{fig:1}.
In Fig.~\ref{fig:3}(a) we show a
blowup of the high density region from Fig.~\ref{fig:2}(b),
indicating more clearly 
that the system forms a triangular lattice which
is commensurate with the underlying square array.
Figure~\ref{fig:2}(c) illustrates
the system in 
Fig.~\ref{fig:1} at $l = 175$ and $d = 1.3$, corresponding
to a local minimum in $M$ and a drop in $P_{6}$.
Although the system still shows 
clustering, the structure of the active disks in the dense region  
is now amorphous, as shown more clearly in
Fig.~\ref{fig:3}(b).
The disorder is produced by a frustration effect
that arises when the natural spacing of the active crystal does not match
the spacing of the interstitial region between the obstacles.
The dip in $M$
is similar to the drop in motion or
decrease in the critical depinning force
found in non-active commensurate 
systems at incommensurate densities
\cite{Reichhardt98a,Berdiyorov06,Bohlein12,Vanossi12,Reichhardt17}.
In the non-active systems,
the commensurate crystalline states
have a higher shear modulus and can be more strongly pinned by the
obstacles.
In contrast, for the frustrated system the shear modulus is reduced,
permitting the particles to move more easily
and producing minima in
the depinning force of the incommensurate state.
Near $d=0.2$ for the active system in Fig.~\ref{fig:1},
a peak in $M$
and a dip in $P_{6}$
appear at another incommensurate region where  
the disks are disordered,
as shown in Fig.~\ref{fig:2}(d) for $d=0.225$.

For $d > 1.5$,
a distinctive type of active cluster appears which has
local square short-range ordering within a single plaquette.
These clusters are associated with a drop in $M$
and an increase in $C$.
An example of the $d=1.7$ clustered state
appears in Fig.~\ref{fig:3}(c),
where a single plaquette  
with local square ordering is highlighted.
As $d$ increases further,
other types of commensurate crystals can occur.  

In the active system near $d = 0.15$,
where there is a smaller peak in $C$, the disks form
a sliding crystalline state where the 
commensuration effect is determined by the number of
rows of active disks that can fit between adjacent rows of obstacles,  
as shown in Fig.~\ref{fig:3}(d) for four rows of disks.
The sliding crystal exhibits intermittent jumping between
crystal and disordered states, causing  
the value of $P_{6}$ to be reduced compared to the
commensurate state which appears at higher $d$.  
If we consider a random array of obstacles,
we do not observe any commensurate effects but instead find a monotonic decrease
of the mobility with increasing $d$. 

\begin{figure}
\includegraphics[width=\columnwidth]{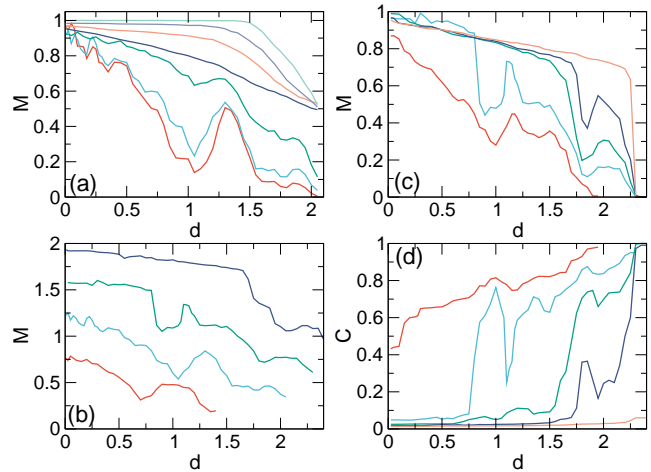}
\caption{ (a) $M$ vs $d$ for the system in
Fig.~\ref{fig:1} for 
$l_{r} = 0.00025$, 0.1, 0.025, 0.25, 40, 175, and $1750$, from top to bottom.
(b) $M$ vs $d$ for $l_{r} = 175$ at
$d_{a}=0.7$, 0.8, 0.9, and $1.0$, from top to bottom,
corresponding to $\phi_{a} = 0.195$, 0.254, 0.32, and $0.4$.
For clarity, the first three curves are shifted up
by $0.915$, 0.61, and $0.305$, respectively, on the $M$ axis.
(c) 
$M$ vs $d$ for samples with
$d_{a} = 0.8$ and $l_{r} = 175$ at varied disk density
$\phi_a=0.00212$, 0.09, 0.17, 0.254, and $0.332$, from top to bottom.  
(d) The corresponding
fraction of particles in the largest cluster $C$
vs $d$ for the system in panel (c). 
}
\label{fig:4}
\end{figure}

In Fig.~\ref{fig:4}(a) we plot $M$ versus $d$ for the system in
Fig.~\ref{fig:1} with fixed $d_a=0.9$ at varied 
$l_{r} = 0.00025$, 0.1, 0.025, 0.25, 40, 175, and $1750$.
The overall mobility decreases with
increasing $l_{r}$,
while commensuration  
effects only appear once $l_{r} > 10$,
which coincides with the running length at which
self-clustering begins to occur.
The commensuration effects become sharper as $l_r$ increases.
For large $d$ and large $l_{r}$, 
the mobility drops to zero
when the system enters an active jammed or clogged state,
while for smaller $l_{r}$ at high $d$,
the flow is reduced but remains finite.  

Figure~\ref{fig:4}(b) shows $M$ versus $d$ for the
system in Fig.~\ref{fig:1} with fixed $l_{r} = 175$
and varied active disk diameter of
$d_{a}= 0.7$, 0.8, 0.9, and $1.0$, giving 
$\phi_{a} = 0.195$, 0.254, 0.32, and $0.4$, respectively.
For $\phi_a = 0.195$, no clustering
occurs until $d > 1.5$, which correlates with a drop 
in $M$ when a commensurate state forms with a structure that
is similar to that illustrated in
Fig.~\ref{fig:3}(c).
The onset of this
locally square commensurate state shifts to lower values of $d$
with increasing $d_{a}$.
The drop in $M$ near $d = 1.0$
in the $\phi_a = 0.254$ system
is due to the
appearance of a different type of
commensurate clustering state where the ordering is triangular rather
than square, similar to what is shown in Fig.~\ref{fig:2}(b) and
Fig.~\ref{fig:3}(a).
The triangular commensurate state persists up to $d =1.1$ 
for the $\phi_a=0.254$ system and appears over a slightly
higher range of $d$
in the $\phi_a=0.32$ system.
The peak near $d = 1.1$ for $\phi_a = 0.254$
is the result of the formation of a frustrated state
of the type illustrated in Fig.~\ref{fig:2}(c) and Fig.~\ref{fig:3}(b).
For $\phi_a = 0.4$, the triangular commensurate state is present
near the dip in $M$ at $d = 0.7$.
In general, as $\phi_a$ increases, the overall magnitude of 
$M$ drops.

The behavior of
$M$ versus $d$
for samples with $l_{r}=175$ where we hold the
active disk diameter 
fixed at $d_{a} = 0.8$ but consider different disk
densities $\phi_{a}=0.00212$, 0.09, 0.17, 0.254, and $0.332$ is shown
in Fig.~\ref{fig:4}(c).
Since the disk radius is fixed,
the
locally square commensuration dip in $M$ at
$d = 1.85$
does not shift with changing $\phi_{a}$.  
For 
$\phi_{a} = 0.00212$, the system is in the single particle limit,
there are no commensurate peaks or dips,
and $M$ drops to zero for $d > 2.25$ when the obstacles form a
percolating barrier to motion.
For $\phi_{a} = 0.09$ and $0.17$,
the
triangular commensuration dip at $d=1.0$ and the
incommensuration peak at
$d=1.1$ found for larger $\phi_a$ 
are absent;
however, there is a high density
incommensuration peak 
at
$d=1.95$.
When $\phi_a = 0.332$, the overall
value of $M$ decreases and additional
incommensuration
peak forms at $d=1.5$.
Here, $M$ drops to zero for $d > 1.9$ when the
system enters an active clogged state.

In Fig.~\ref{fig:4}(d) we plot the largest cluster size $C$
versus $d$ for the system in Fig.~\ref{fig:4}(c) with
varied $\phi_a$.
When
$\phi_a = 0.00212$, $C$ remains small indicating the lack of any clustering
in the single particle limit, 
while for $\phi_a = 0.09$ and $\phi_a=0.18$,
$C$ approaches $C=1.0$ 
for $d > 2.25$ and has a peak at $d = 1.8$
corresponding to the formation of local square ordering in individual
substrate plaquettes.
At $\phi_a = 0.254$, there are three peaks
in $C$ corresponding to the commensuration
effects at $d = 1.0$, 1.35, and $1.85$, as well as
a dip produced by a frustrated state at $d = 1.1$.
For $\phi_a = 0.332$, a clustered state appears for all values
of $d$
which develops crystalline ordering at the three
commensuration values of $d$.

{\it Summary---}
We have examined run-and-tumble active matter disks interacting
with a periodic obstacle array and find 
novel active matter commensuration and frustration effects.
These arise
when the active matter
undergoes motility-induced phase separation into
a dense crystalline phase which has a natural disk spacing.
When this spacing is commensurate with the lattice constant of the obstacle
array,
a large crystalline phase separated
state can appear, whereas
for other obstacle spacings,
the crystalline phase cannot fit on the substrate and we instead find
a frustrated state
in which the clusters are amorphous and not as large.
The commensuration and incommensuration effects
produce
peaks and dips in the mobility, six-fold order, and cluster size
as a function of changing obstacle diameter.
The commensurate
crystal states can have long range triangular
ordering or local square ordering. At low activity
or in the Brownian limit, the commensuration effects are lost.    

\begin{acknowledgments}
We gratefully acknowledge the support of the U.S. Department of
Energy through the LANL/LDRD program for this work.
This work was supported by the US Department of Energy through
the Los Alamos National Laboratory.  Los Alamos National Laboratory is
operated by Triad National Security, LLC, for the National Nuclear Security
Administration of the U. S. Department of Energy (Contract No. 892333218NCA000001).
\end{acknowledgments}

\bibliography{mybib}

\end{document}